\documentclass[paper=letter, fontsize=12pt]{article}
\usepackage[english]{babel} 
\usepackage{amsmath,amsfonts,amsthm} 
\usepackage[utf8]{inputenc}
\usepackage{float}
\usepackage{lipsum} 
\usepackage{blindtext}
\usepackage{graphicx}
\usepackage{caption}
\usepackage{subcaption}
\usepackage[sc]{mathpazo} 
\usepackage[T1]{fontenc} 
\usepackage{microtype} 
\usepackage[hmarginratio=1:1,top=32mm,columnsep=20pt]{geometry} 
\usepackage{multicol} 
\usepackage{booktabs} 
\usepackage{float} 
\usepackage{hyperref} 
\usepackage{bookmark}
\usepackage{lettrine} 
\usepackage{paralist} 
\usepackage{abstract} 
\usepackage{titlesec} 

\renewcommand\thesection{\Roman{section}} 
\renewcommand\thesubsection{\Roman{subsection}} 

\titleformat{\section}[block]{\large\scshape\centering}{\thesection.}{1em}{} 
\titleformat{\subsection}[block]{\large}{\thesubsection.}{1em}{} 
\usepackage{fancyhdr} 
\newcommand{\ud}{\mathrm{d}}

\title{\vspace{-15mm}\fontsize{24pt}{10pt}\selectfont\textbf{Distributional Mellin calculus in $\mathbb{C}^n$, with applications to option pricing}} 
\author{
{Jean-Philippe Aguilar$^{a,\dagger}$, Cyril Coste$^{b,\dagger}$, Hagen Kleinert$^{c,\dagger\dagger}$, Jan Korbel$^{d,\dagger\dagger\dagger}$ }\\[2mm]
{\it $\dagger$ BRED Banque Populaire, Modeling Department, 18 quai de la Râpée, Paris - 75012}\\[2mm]
{\it $\dagger\dagger$ Institute for Theoretical Physics, Freie Universit\"at Berlin, Arnimallee 14, 14195, Berlin}\\[2mm]
{and {\it  ICRANeT Piazzale della Repubblica, 10 -65122, Pescara}}\\[2mm]
{\it $\dagger\dagger\dagger$ Department of Physics, Zhejiang University, Hangzhou 310027, PRC}\\[2mm]
{and {\it  Faculty of Nuclear Sciences and Physical engineeering, B\v{r}ehov\'{a} 7, Prague - 11519}}\\[2mm]
{\textit{(a)} jean-philippe.aguilar@bred.fr, \textit{(b)} cyril.coste@bred.fr,}\\[2mm]
{(c) h.k@fu-berlin.de, (d) korbeja2@fjfi.cvut.cz}
}

\date{}

\providecommand{\keywords}[1]{\textbf{\textit{Key words---}} #1}
\newcommand{\res}{\mathrm{Res}}

\begin{document}
\maketitle 
\tableofcontents
\pagestyle{headings}
\setcounter{page}{1}
\pagenumbering{arabic}
\begin{abstract}
\noindent We discuss several aspects of Mellin transform, including distributional Mellin transform and inversion of multiple Mellin-Barnes integrals in $\mathbb{C}^{n}$ and its connection to residue expansion or evaluation of Laplace integrals. These mathematical concepts are demonstrated on several option-pricing models. This includes European option models such as Black-Scholes or fractional-diffusion models, as well as evaluation of quantities related to the optimal exercise price of American options. 
\end{abstract}

\thispagestyle{fancy} 

\section{Introduction}
Mellin transform belongs, together with Fourier transform, Laplace transform and Z-transform, to the most important integral transforms with many applications in applied mathematics, theory of diffusion, quantum mechanics, image recognition, statistics or finance. It can be interpreted as a multiplicative version of two-sided Laplace transform and therefore it can be successfully used for description of multiplicative processes and other diffusion-like phenomena as well as for description and analysis of various special functions - Mittag-Leffler functions, Fox-H function \cite{Gorenflo99,Mainardi05}, stable distributions \cite{Kleinert16} or option pricing , just to name a few. Within the past few years, the one-dimensional Mellin transform has also been applied to various option pricing problems \cite{Pagnini04,Frontczak11}. Although it is an integral transformation in the complex domain, it can be often successfully used in numerical calculations and asymptotic formulas, in particular in analytic number theory \cite{Flajolet95} or in quantum field theory \cite{Aguilar08}. 

\noindent Nevertheless, some aspects of Mellin transform and Mellin-Barnes integral representation have not been satisfactorily discussed. First, generalization of Mellin transform to the domain of generalized functions enables to deal with wider class of functions, including polynomials and logarithmic functions. As a byproduct, we obtain a generalized version of delta function, which is, in analogy with Fourier transform, Mellin transform of a constant; within this framework, we are therefore able to investigate various functions, including powers or powers of logarithms. We call this new approach \textit{"distributional Mellin calculus"}.

\noindent Second, generalization of Mellin calculus to two or more dimensions and appropriate techniques for its calculation represents a conceptually new powerful tool for analysis of asymptotic properties of many systems, based on theory of residues in $\mathbb{C}^n$ and residue representation. This enables us to perform the complex integrals and end with a (Grothendieck) residue series, which can be used for numerical calculations as well as for description of asymptotic properties. This approach, first formalized in \cite{Tsikh94,Tsikh97} (without distributional aspects) was already used in \cite{AGDR08} to estimate analytically some standard model contributions to the muon anomalous moment.

\noindent The new and powerful combination of distributional Mellin calculus and residue theory in $\mathbb{C}^n$ can be successfully used in many applications, for instance in generalized diffusion or option pricing \cite{AC16}. In this paper we make more examples, including Gaussian and non-Gaussian models as well as an application to the evaluation of Laplace integrals. These integrals appear in various applications as, for instance, the optimal exercise price of an American option \cite{Zhu06}. 

\noindent The paper is organized as follows: section 2 discusses basic properties of Mellin transform and Mellin-Barnes integrals and generalizes it to distributional calculus. As an example, Mellin-Barnes representation of European call option for Black-Scholes model and space-time fractional model is presented. Section 3 introduces n-dimensional Mellin calculus and presents residue theorem for Mellin-Barnes integral. Consequently, series formula for European Black-Scholes call option is calculated. Section 4 discusses applications of multidimensional Mellin calculus to evaluation of Laplace integrals and its application to American options. Section 5 is devoted to conclusions and perspectives.

\section{Distribution theory of the one-dimensional Mellin transform}
\label{sec:Distribution}

\subsection{Definition and properties}
See \cite{Flajolet95} for an excellent overview of the Mellin transform and its applications, and \cite{Erdélyi54} for a dictionnary of frequently used transforms.
\newline
The Mellin transform of a locally integrable function $f$ on $(0,\infty)$ is the function $f^*:\mathbb{C}\rightarrow\mathbb{C}$ defined by
\begin{equation}\label{mellin_def}
f^*(z) \, = \, \int\limits_0^\infty f(x) x^{z-1}\, dx
\end{equation}
In general, the integral (\ref{mellin_def}) converges on a \textit{strip} (the so-called \textit{fundamental strip} of $f^*$) in $\mathbb{C}$, that is an open subset of the type
\begin{equation}
\langle a , b \rangle \, = \, \{z\in\mathbb{C}, Re(z) \in (a,b) \}
\end{equation}
where $a$ and $b$ are determined by the asymptotic behaviour of $f$ around $0$ and $\infty$. In this framework, the Gamma function, also known as \textit{Euler's integral of the second kind} and
defined by
\begin{equation}\label{gamma_function}
\Gamma(z) \, = \, \int\limits_0^\infty e^{-x} x^{z-1} \, dx
\end{equation}
is the Mellin transform of $e^{-x}$ with fundamental strip $\langle 0,\infty \rangle$. It can be extended to the left-half plane thanks to the functional relation
\begin{equation}\label{functional_gamma}
z(z+1) \cdots (z+n) \Gamma (z) \, = \, \Gamma(z+n) \,\,\, , \,\,\,\,\, n \in \mathbb{N}
\end{equation}
unless for negative integers $-n$ where (\ref{functional_gamma}) shows that $\Gamma(z)$ possesses a pole of residue $\frac{(-1)^n}{n!}$. We will express this fact by writing
\begin{equation}\label{singular_series_gamma}
\Gamma(z) \, \asymp \, \sum\limits_{n=0}^{+\infty}
\frac{(-1)^n}{n!}\frac{1}{z+n}
\end{equation}
where the formal series in the right hand side of (\ref{singular_series_gamma}) is called the \textit{singular series} of the Gamma function.

\noindent Another important Mellin transform follows from the particular case of \textit{Beta integral}:
\begin{equation}
\int\limits_0^\infty \frac{x^{z-1}}{1+x} \, dx \, = \, \mathrm{B}(z,1-z) \, = \, \Gamma(z) \Gamma(1-z)
\end{equation}
which means that $\Gamma(z) \Gamma(1-z)$ is the Mellin transform of $\frac{1}{1+x}$ with fundamental strip $\langle 0 , 1\rangle$. Its singular series follows easily from (\ref{singular_series_gamma}) and from the property $\Gamma(1+n) \, = \, n!$ for $n\in\mathbb{N}$:
\begin{equation}\label{singular_series_beta}
\Gamma(z)\Gamma(1-z) \, \asymp \, \sum\limits_{n=-\infty}^{+\infty}
(-1)^n\frac{1}{z+n}
\end{equation}
In Table 1, we list some important properties of the Mellin Transform which follow directly from definition (\ref{mellin_def}).
\begin{table}[h]
\label{mellin_properties}
\centering
\begin{tabular}{l l l l }
\hline
$\mathbf{f(x)}$ & $\mathbf{f^*(z)}$ & \textbf{Fundamental strip}  & \\
\hline
$f(x^\alpha)$  & $\frac{1}{\alpha}f^*(\frac{s}{\alpha})$ & $\langle \alpha a , \alpha b \rangle$ & $\alpha > 0$ \\
$x^\beta f(x)$  & $f^*(\beta +s)$ & $\langle a - \beta ,  b - \beta \rangle $ & \\
$f(1/x)$  & $f^*(-s)$ & $\langle -b ,  -a \rangle $ & \\
$f(\mu x)$  & $\frac{1}{\mu^s}f^*(s)$ & $\langle a , b \rangle$ & $\mu > 0 \,\,\,  \textit{['Rescaling']}$ \\
$\frac{d}{dx}f(x)$  & $-(z-1)f^*(z-1)$  & $\langle a -1 , b -1 \rangle$ & \\$f(x) \log ^n x$  & $\frac{d^n}{dz^n}f^*(z)$ & $\langle a  , b  \rangle$ & \\
\hline
\end{tabular}
\caption{Basic properties of the Mellin Transform}
\end{table}

\noindent \noindent \textbf{\underline{Example (Heat kernel):}} Let $g(x,t,\sigma)$ denote the heat kernel:
\begin{equation}
g(y,t,\sigma) \, = \, \frac{1}{\sigma\sqrt{2\pi}\tau} \, e^{-\frac{y^2}{2\sigma^2 \tau}}
\end{equation}
Using the properties in table 1, it is immediate to see that its $\sigma$-Mellin transform writes
\begin{equation}\label{mellin_heat}
g^*(y,t,z) \, = \, \frac{1}{\sqrt{2\pi}} \times  \frac{1}{2} \Gamma \left(-\frac{z}{2} \right) \left( \frac{y}{\sqrt{2}}  \right)^{z} (\sigma\sqrt{\tau})^{-(1+z)}
\end{equation}

\subsection{Inverse Mellin-Barnes integrals}

The inversion of the Mellin transform is performed via an integral along any vertical line in the fundamental strip:
\begin{equation}\label{Mellin_inversion}
f(x) \, = \, \frac{1}{2i\pi} \int\limits_{\gamma-i\infty}^{\gamma+i\infty} f^*(z) \, x^{-z} \, dz \,\,\, , \,\, \gamma\in (a,b)
\end{equation}
and in the case where $f^*$ is a ratio of products of Gamma functions of linear argument, integral (\ref{Mellin_inversion}) is said to be a \textit{Mellin-Barnes integral}:
\begin{equation}\label{Mellin-Barnes}
f(x) \, = \, \frac{1}{2i\pi} \int\limits_{\gamma-i\infty}^{\gamma+i\infty}
\frac{\prod_j \Gamma(a_j z + b_j)}{\prod_k \Gamma(c_k z + d_k)} x^{-z} \, dz \, \, , \,\,\, a_j, b_j, c_k, d_k \in \mathbb{R}
\end{equation}
The question at stake is whether (\ref{Mellin-Barnes}) can be expressed as a sum of residues of the integrand in some region of $\mathbb{C}$, that is, if we can close the vertical line of integration in such a way that the integrand does not contribute in this part of the contour when $|z|\rightarrow\infty$. To answer this question, introduce the characteristic quantity
\begin{equation}\label{Delta_1}
\Delta \, := \, \sum_j a_j \, - \, \sum_k c_k
\end{equation}
Using the Stirling approximation for the Gamma function, it can be shown that the integrand in (\ref{Mellin-Barnes}) decreases exponentially in the half-plane (see \cite{Tsikh94} for $\Delta \neq 0$ and \cite{Tsikh97} for $\Delta = 0$):
\begin{equation}\label{Pi_Delta_1}
\Pi_\Delta \, := \, \{ z \in \mathbb{C}, \, Re(\Delta z) < \Delta.\gamma \}
\end{equation}
If $\Delta\neq 0$, $\Pi_\Delta$ then reduces to:
\begin{align}\label{BS_Formula}
\left\{
\begin{aligned}
 & \textrm{If }\Delta > 0 \,\,  \textrm{then} & \Pi_\Delta = \Pi^- := \{ z \in \mathbb{C}, \, Re(z) < \gamma \} \\
 & \textrm{If }\Delta < 0 \,\,  \textrm{then} & \Pi_\Delta = \Pi^+ := \{ z \in \mathbb{C}, \, Re(z) > \gamma \}
\end{aligned}
\right.
\end{align}
which means that, depending on the sign of $\Delta$, (\ref{Mellin-Barnes}) can be expressed as a sum of residues in one half-plane (left to the fundamental strip if $\Delta > 0$, right if $\Delta <0$). If $\Delta = 0$, residues summations hold in both half-planes.

\textbf{\underline{Example 1.}} Let $f(x) = e^{-x}$, then $f^*(z)= \Gamma(z)$ with fundamental strip $\langle 0, \infty \rangle$, $\Delta = +1$ and therefore
\begin{equation}\label{exp_series}
 f(x) \, = \, \sum\limits_{Re(z_n) < 0} \res_{z_n} \Gamma(z)x^{-z}
 \, = \, \sum\limits_{n=0}^{\infty} \frac{(-1)^n}{n!}x^n
\end{equation}
where for the last equality we have used the singular series (\ref{singular_series_gamma}) for the Gamma function; note (\ref{exp_series}) is the Taylor expansion of $e^{-x}$ around $0$.

\textbf{\underline{Example 2.}} Let $f(x) = \frac{1}{1+x}$, then $f^*(z) = \Gamma(z) \Gamma(1-z)$ with fundamental strip $\langle 0,1 \rangle$ and $\Delta = 0$.
One can either left sum the residues of $f^*(z)$:
\begin{equation}
f(x) \, = \, \sum\limits_{Re(z_n) < 0} \res_{z_n} \Gamma(z)\Gamma(1-z) x^{-z}
 \, = \, \sum\limits_{n=0}^{\infty} (-1)^n x^n
\end{equation}
which is the series expansion of $\frac{1}{1+x}$ for small x, either right sum the residues:
\begin{equation}
f(x) \, = \, \sum\limits_{Re(z_n) > 0} \res_{z_n} \Gamma(z)\Gamma(1-z) x^{-z}
 \, = \, \sum\limits_{n=0}^{\infty} (-1)^n x^{-(1+n)}
\end{equation}
which is the series expansion of $\frac{1}{1+x}$ for large x. Both residues are directly computed from the singular series (\ref{singular_series_beta}).

\subsection{Generalized Mellin transform}

The classical Mellin theory does not take into account several elementary functions such as powers or powers of logarithms, because their Mellin transform (\ref{mellin_def}) has an empty fundamental strip. For instance, by definition, the Mellin transform of $x^\alpha$ is the integral
\begin{equation}\label{xalpha}
\int\limits_0^\infty x^{\alpha + z - 1}\,dx
\end{equation}
which converge for no value of $z$. But, performing the change of variables $x=e^y$, the integral (\ref{xalpha}) becomes:
\begin{equation}
\int\limits_{-\infty}^{+\infty} e^{(\alpha + z)y}\,dy \, = \, \int\limits_{-\infty}^{+\infty} e^{2i\pi ( \frac{\alpha + z}{2i\pi} ) y}\,dy \, = \, \delta_{( \frac{\alpha + z}{2i\pi} )}
\end{equation}
where we have used the fact that the Fourier transform of $1$ is the Dirac distribution $\delta$ (see \cite{Gelfand64}). Note that a formal application of the inversion formula (\ref{Mellin_inversion}) and a change of variables $z = 2i\pi z'$ yields
\begin{equation}
\frac{1}{2i\pi} \int\limits_{-i\infty}^{+i\infty} \delta_{( \frac{\alpha + z}{2i\pi} )} x^{-z} \, dz \, = \, \left\langle  \delta_{( \frac{\alpha}{2i\pi} + z' )} \, , \, x^{-2i\pi z '} \right\rangle \, = \, x^{\alpha}
\end{equation}
where the braket stands for the distributional duality. In particular, the Mellin transform of $x^0=1$ is $\delta_{( \frac{z}{2i\pi})}$ and using the "logarithms" rule in table 1, we deduce that the Mellin transform of $\log ^n x$ is
\begin{equation}
\frac{d^n}{dz^n} \delta_{( \frac{z}{2i\pi})} \, = \,  \frac{1}{(2i\pi)^n} \delta^{(n)}_{( \frac{z}{2i\pi})}
\end{equation}
This fact can be recovered by considering the integrals:
\begin{align}
\Omega_n(z) & = \int\limits_0 ^\infty \log ^n x \, x^{z-1} \, dx \\
& = \int\limits_{-\infty}^{+\infty} e^{2i\pi ( \frac{z}{2i\pi} ) y} \, y^n \, dy \, = \, \frac{1}{(2i\pi)^n} \delta^{(n)}_{( \frac{z}{2i\pi})}
\end{align}
(The Fourier transform of integer powers results in derivatives of the Dirac distribution). In table 2 we summarize some useful Mellin transforms, obtained either by the "classical" theory or the distributional one.
\begin{table}[h]
\label{mellin_examples}
\centering
\begin{tabular}{l l l }
\hline
$\mathbf{f(x)}$ & $\mathbf{f^*(z)}$ & \textbf{Fundamental strip}   \\
\hline
$e^{-x}$ & $\Gamma(z)$ & $ \langle 0,\infty \rangle $ \\
$\frac{1}{1+x}$ & $\Gamma(z)\Gamma(1-z)$ & $ \langle 0,1 \rangle $ \\
\vspace*{0.2cm}
$x^\alpha$ & $ \delta_{( \frac{\alpha + z}{2i\pi})}  $ & $ \emptyset $ \\
\vspace*{0.2cm}
$1$ & $ \Omega_0(z)\, = \, \delta_{( \frac{z}{2i\pi})}  $ & $ \emptyset $ \\
$\log^n x$ & $ \Omega_n(z)\, = \, \frac{1}{(2i\pi)^n}\delta^{(n)}_{( \frac{z}{2i\pi})}  $ & $ \emptyset $ \\
\hline
\end{tabular}
\caption{Some useful Mellin pairs}
\end{table}

\noindent The particular values we have obtained for $\Omega_n(z)$ express the fact that the distribution-valued complex function $p\rightarrow\Omega_p(z)$ is analytic in the right complex plane (as the Fourier transform of a so-called \textit{homogeneous distribution} \cite{Gelfand64,Franssens11}). One can in fact meromorphically extend $\Omega_p(z)$ to negative values of $p$ because a simple integration by parts shows that the following functional relation holds:
\begin{equation}
\frac{\Omega_p(z)}{z} \, = \, - \frac{\Omega_{p+1}(z)}{p+1}
\end{equation}
which means that $\frac{\Omega_p(z)}{z}$ possesses a pole in $p=-1$ whose residue is $-\Omega_0(z) = - \delta_{( \frac{z}{2i\pi})}$. Iterating the procedure yields:
\begin{equation}\label{singular_series_omega}
\frac{\Omega_p(z)}{z^n} \, = \, \frac{(-1)^{n}\Omega_{p+n}(z)}{(p+1)(p+2)\cdots(p+n)} \underset{p\rightarrow -n}{\sim} -\frac{1}{(n-1)!}\frac{\delta_{( \frac{z}{2i\pi})}}{p+n}
\end{equation}
which can be regarded as the distributional analogue to the singular series of the Gamma function (\ref{singular_series_gamma})

\subsection{The Mellin-Barnes representation for an European Black-Scholes call}

Recall that the Black-Scholes price of an European call option can be written as \cite{Wilmott06}:
\begin{equation}\label{BS_Green}
V_{BS}(S,K,\tau) \, = \, e^{-r\tau} \int\limits_{-\infty}^{+\infty}  \underbrace{[Se^{(r-\frac{1}{2}\sigma^2)\tau + y}-K]^+}_{\textrm{"Modified" payoff}} \, \times \, \underbrace{g(y,\tau,\sigma)}_{\textrm{Heat kernel}} dy
\end{equation}
It is easy to compute the Mellin transform of the payoff \cite{Manuge13,AC16} and therefore obtain the representation:
\begin{equation}
\label{payoffmellin}
[Se^{(r-\frac{1}{2}\sigma^2)\tau + y}-K]^+ \, = \, \frac{K}{2i\pi} \int\limits_{c_s-i\infty}^{c_s+i\infty} -\frac{e^{-(r-\frac{1}{2}\sigma^2)\tau s -ys}}{s(s+1)}\Big(\frac{S}{K}\Big)^{-s}\, ds \, \, \, , \, \, \, \, c_s\in (-\infty, -1)
\end{equation}
Introducing the Mellin-Barnes representation for the heat kernel (\ref{mellin_heat}) and a similar representation for $e^{\frac{1}{2}\sigma^2\tau s}$ (which follows directly from the properties in table 1) gives birth to two more integrals, over complex variables we shall denote by $t_1$ and $t_2$. We are then left with the $y$-integral:
\begin{equation}
\int\limits_{-\infty}^{+\infty} e^{-ys} \, y^{t_1} \,dy ^, = \, \Omega_{t_1} (-s)
\end{equation}
Replacing in (\ref{BS_Green}), we have:
\begin{multline}\label{BS_Green2}
V_{BS}(S,K,\tau) \, = \, \frac{1}{(2i\pi)^3}\frac{Ke^{-r\tau}}{4\sqrt{2\pi}} \int\limits_{c_s-i\infty}^{c_s+i\infty} \int\limits_{\gamma_1-i\infty}^{\gamma_1+i\infty} \int\limits_{\gamma_2-i\infty}^{\gamma_2+i\infty}
-(-1)^{-\frac{t_2}{2}}2^{\frac{t_2-t_1}{2}} \, \times \\
\frac{s^{-\frac{t_2}{2}}\Omega_{t_1}(-s)e^{-s(\log\frac{S}{K}+r\tau)}}{s(s+1)} 
\times \Gamma(-\frac{t_1}{2})\Gamma(\frac{t_2}{2}) (\sigma\sqrt{\tau})^{-(1+t_1+t_2)} \, ds \wedge dt_1 \wedge dt_2 \\
\end{multline}
where $\gamma_1<0$ (resp. $\gamma_2>0$), corresponding to the fundamental strip of the $\Gamma(-\frac{t_1}{2})$ (resp. $\Gamma(\frac{t_2}{2})$) function. Now, introducing the vector $\underline{\gamma}:=(\gamma_1,\gamma_2)$, denoting
\begin{equation}
\underline{\gamma} \, + \, i\mathbb{R}^2 \, := \, (\gamma_1-i\infty , \gamma_1+i\infty) \, \times \, (\gamma_2-i\infty , \gamma_2+i\infty)
\end{equation}
and using the change of variables $s\rightarrow 2i\pi s$, (\ref{BS_Green2}) becomes:
\begin{equation}\label{BS_Green3}
V_{BS}(S,K,\tau) \, = \,
\frac{1}{(2i\pi)^2}\frac{Ke^{-r\tau}}{4\sqrt{2\pi}} \int\limits_{\underline{\gamma}+i\mathbb{R}^2}
2^{\frac{t_2-t_1}{2}}
N(t_1,t_2)
\Gamma(-\frac{t_1}{2})\Gamma(\frac{t_2}{2}) (\sigma\sqrt{\tau})^{-(1+t_1+t_2)} \, d\underline{t}
\end{equation}
where $d\underline{t}:=dt_1 \wedge dt_2$, and where $N(t_1,t_2)$ is defined by the action of the distribution $\Omega_{t_1}(-2i\pi s)$ over a certain test function:
\begin{equation}
N(t_1,t_2) \, = \, (-2i\pi)^{-(1+\frac{t_2}{2})} \, \left\langle \,   \frac{s^{-\frac{t_2}{2}}}{s(1 + 2i\pi s)}\Omega_{t_1}(-2i\pi s) \, , \,  e^{-2i\pi s \left(\log\frac{S}{K}+r\tau \right)}  \, \right\rangle
\end{equation}

To perform the integral (\ref{BS_Green3}), we need to extend the concepts we have introduced in this section to the case of $\mathbb{C}^n$, $n>1$; this will be the object of the next section.

\subsection{The Mellin-Barnes representation for an option driven by fractional diffusion}

Black-Scholes formula has been very popular among practitioners, nevertheless its applicability is limited due to simplified assumptions excluding complex phenomena as sudden jumps, memory effects, etc. There have been studied various generalizations of Black-Scholes formula, let us mention regime switching multifractal models~\cite{Calvet08}, stochastic volatility models~\cite{Heston93} or jump processes~\cite{Tankov03}. Particularly interesting are generalized diffusion models, especially models with fractional diffusion. These models are also closely related to Mellin transform. Space-time (double)-fractional diffusion equation can be expressed as
\begin{equation}\label{double_fractional}
\left({}^\ast_0 \mathcal{D}^\gamma_t + \mu  [{}^\theta \mathcal{D}^\alpha_x] \right) g(x,t) = 0\, , \ \mu > 0
\end{equation}
where ${}^\ast_0 \mathcal{D}^\gamma_t$ is the \emph{Caputo fractional derivative} and ${}^\theta \mathcal{D}^\alpha_x$ is the \emph{Riesz-Feller fractional derivative}. All technical details and a broad mathematical discussion can be found in \cite{Gorenflo99}. Eq. \eqref{double_fractional} can be solved by in terms of Mellin-Barnes integral as

\begin{equation}\label{Mellin_Barnes_DF}
g_{\alpha,\gamma}^\theta(x,t,\mu) = \frac{ 1 }{\alpha x}\frac{1}{2 \pi i}   \int\limits_{c_1 - i \infty}^{c_1 + i \infty}
\Gamma \left[
         \begin{array}{ccc}
           \frac{t_1}{\alpha} & 1-\frac{t_1}{\alpha} & 1-t_1 \\
           1-\frac{ \gamma s}{\alpha} & \frac{\alpha-\theta}{2\alpha}t_1 & 1 - \frac{\alpha-\theta}{2\alpha}t_1 \\
         \end{array}
       \right] \left(\frac{x}{(-\mu t^{\gamma})^{1/\alpha}}\right)^{t_1} \, dt_1
\end{equation}
where \emph{Gamma fraction} is defined as $\Gamma\left[\begin{array}{ccc}
                                                         x_1 & \dots & x_n \\
                                                         y_1 & \dots & y_m
                                                       \end{array}\right] = \frac{\Gamma(x_1)\dots\Gamma(x_n)}{\Gamma(y_1)\dots\Gamma(y_m)}$

\noindent The corresponding European call option price has been calculated in \cite{Kleinert16} as:
\begin{equation}
V_{(\alpha,\gamma,\theta)}(S,K,\tau) = e^{-r \tau} \int\limits_{-\infty}^\infty  \ \left[S e^{\tau(r-q+\mu)+y} - K\right]^+ g_{\alpha,\gamma}^\theta(y,\tau,\mu) \ud y\, 
\end{equation}
which is a generalization of the Black-Scholes propagator (\ref{BS_Green}). Parameter $\mu$ appearing the ``modified payoff'' has its origin in the risk-neutral (or martingale-equivalent) probability measure, which is for exponential transform obtained by so-called \emph{Esscher transform} \cite{Gerber93}. It can be calculated as
\begin{equation}
\mu = \ln \int e^y g_{\alpha,\gamma}^\theta(y,\tau=1) \ud y\, 
\end{equation}
Unfortunately, it is not always possible to express $\mu$ analytically. Nevertheless in the degenerate Gaussian case (that is for $\alpha=2$ and $\gamma=1$) , we have $\mu = -\frac{1}{2}\sigma^2$, and the fractional derivatives in (\ref{double_fractional}) reduce to the usual heat equation.

\noindent By very similar manipulations to the ones we have made in the Black-Scholes case, it is possible to derive the Mellin-Barnes representation for European call option driven by space-time fractional diffusion as
\begin{multline}\label{MB_DF}
V_{(\alpha,\gamma,\theta)}(S,K,\tau) = \frac{K e^{-r \tau}}{\alpha (2\pi i)^2} \times \\ 
\int\limits_{\underline{c}+i\mathbb{R}^2} N(t_1,t_2) \Gamma \left[
         \begin{array}{ccc}
           \frac{t_1}{\alpha} & 1-\frac{t_1}{\alpha} & 1-t_1 \\
           1-\frac{ \gamma t_1}{\alpha} & \frac{\alpha-\theta}{2\alpha}t_1 & 1 - \frac{\alpha-\theta}{2\alpha}t_1 \\
         \end{array}
       \right]
       \Gamma(t_2) 
       \left(-\mu \tau\right)^{-(\frac{t_1}{\alpha}+t_2)}.\tau^{\frac{1-\gamma}{\alpha}t_1}
       \,  d\underline{t}
\end{multline}
where
\begin{equation}
N(t_1,t_2) = (-2\pi i)^{-(1+t_2)} \left\langle \frac{s^{-t_2}}{s(1+ 2\pi i s)}\Omega_{t_1-1}(- 2 \pi i s), e^{- 2 \pi i s (\log \frac{S}{K} + r \tau)} \right\rangle\, 
\end{equation}
Let us note that this representation slightly differs from the representation of Black-Scholes model, because here we have used slightly different Mellin-Barnes representation of Green function $g(y,\tau)$, which for Black-Scholes model (Gaussian distribution) is equal to

\begin{equation}\label{Heat_Kernel_mod}
g(y,\tau,\sigma) = \frac{1}{2 \sigma \tau^{1/2} y} \frac{1}{2\pi i} \int\limits_{c_1-i \infty}^{c_1+i \infty} \frac{\Gamma(1-t_1)}{\Gamma(1-t_1/2)} \left(\frac{y}{\sigma \tau^{1/2}}\right)^{t_1} dt_1 
\end{equation}
Of course (\ref{Heat_Kernel_mod}) is equivalent to the Mellin-Barnes representation (\ref{mellin_heat}) for the heat kernel: it follows easily from the change of variables $t_1'=t_1-1$ and an application of the Legendre duplication formula for the Gamma function. Last, note that the extra $\tau^{\frac{1-\gamma}{\alpha}t_1}$ term is induced by the fractional nature of the time derivatives and disappears when $\gamma = 1$ (natural derivatives). The analytic evaluation of the representation (\ref{MB_DF}) will be the subject of a future research article.

\section{Inversion of muliple Mellin-Barnes integrals}
\label{sec:invmultipleMB}

\subsection{An introduction to the calculus of residues in $\mathbb{C}^n$}
This brief presentation of fundamental $\mathbb{C}^n$ tools is directly inspired by the classical book \cite{Griffiths78}, as well as the doctoral dissertation \cite{Aguilar08}.

\subsubsection{Complex differential $n$-forms}

Let $n>1$ an integer, and denote
\begin{equation}
z := (z_1, \dots \, z_n) \in \mathbb{C}^n \,\,\, , \,\,\,\, \overline{z} := (\overline{z_1}, \dots \, \overline{z_n}) \in \mathbb{C}^n
\end{equation}
and
\begin{equation}
dz := dz_1 \wedge \dots \wedge dz_n  \,\,\, , \,\,\,\, d\overline{z} := d\overline{z_1} \wedge \dots \wedge d\overline{z_n}
\end{equation}
where we use the usual wedge "$\wedge$" to symbolize the properties
\begin{align}\label{wedge}
\left\{
\begin{aligned}
& dx \wedge dy \, = \, -dy \wedge dx \\
& dx \wedge dx \, = \, 0
\end{aligned}
\right.
\end{align}
The $2n$-vector $(dz,d\overline{z})$ forms a basis of the cotangent space $T_x^*\mathbb{C}^n$, meaning that the total differential of an application $f:U\subset\mathbb{C}^n \rightarrow \mathbb{C}$ can be written under the form
\begin{equation}
df \, = \, \partial f dz \, + \, \overline{\partial} f d\overline{z}
\end{equation}
We will say that $f$ is \textit{holomorphic} in the open set $U\subset\mathbb{C}^n$, and denote $f\in\mathcal{O}_n(U)$, if the Cauchy-Riemann equation
\begin{equation}\label{cauchy_riemann}
\overline{\partial} f \, = \, 0
\end{equation}
is satisfied in $U$. In particular, if $U=\{0_{\mathbb{C}^n}\}$, we will denote by $\mathcal{O}_n$ the \textit{ring of holomorphic functions in $0=0_{\mathbb{C}^n}$}. Now, let us introduce complex differential $n$-forms on $U$, that is, elements of $T_x^*U$ which have the generic form
\begin{equation}
\omega \, = \, f \, dz \, + \, g \, d\overline{z}
\end{equation}
for some functions $f$ and $g$ on $U$. If in particular $g=0$, then by properties (\ref{wedge}) the differential of $\omega$ writes
\begin{equation}
\begin{aligned}
d\omega & =\,  (\partial f dz \, + \, \overline{\partial} f d\overline{z}) \, \wedge \, dz \\
& = \, -\overline{\partial}f \, dz \wedge d \overline{z}
\end{aligned}
\end{equation}
If $f$ is holomorphic in $U$, then by definition (\ref{cauchy_riemann}) we have $\overline{\partial} f = 0$ and thus $d\omega \, = \, 0$ ($\omega$ is closed). Therefore, for any (orientable) subset $\Omega\subset U$ we have, by Stokes' theorem
\begin{equation}
\int\limits_{\partial\Omega} \, \omega \, = \, \int\limits_\Omega \, d\omega \, = \, 0
\end{equation}
This constitutes the $\mathbb{C}^n$ analogue to the Cauchy integral theorem in $\mathbb{C}$.

\subsubsection{Grothendieck residue}

Let $f_1, \dots ,f_n\in\mathcal{O}_n$ a sequence in $\mathcal{O}_n$; it is called a \textit{regular sequence} if, for any $j=1 \dots n$, $f_j$, $f_j$ is a non-zero divisor in the quotient ring $\mathcal{O}_n / (f_1,\dots f_{j-1})$, where $(f_1,\dots f_{j-1})$ is the ideal generated by $f_1 \dots f_{j-1}$.

\textbf{\underline{Example.}} Let $n=2$; then the sequence $(z_1,z_2)$ is a regular sequence, but $(z_1,z_1z_2)$ is not, because the first term of the sequence divides the second one. More generally, the elements of a regular sequence in $\mathcal{O}_n$ all vanish in $0_{\mathbb{C}^n}$; this comes from the fact that $\mathcal{O}_n$ is a so-called \textit{local} ring, that is, possesses a unique maximal ideal, and it can be shown that this ideal is generated by $(z_1, \dots ,z_n)$.
Now, consider applications $g,f_1,\dots,f_n$ in $\mathcal{O}_n$ and suppose that $f_1,\dots,f_n$ is a regular sequence; let $\omega$ be the differential form
\begin{equation}
\omega \, = \, \frac{g}{f_1 \dots f_n} \, dz
\end{equation}
Then, the quantity
\begin{equation}
\res_0 \, := \, \frac{1}{(2i\pi)^n}\int\limits_{\substack{|z_1|=\epsilon_1\\ \dots \\ |z_n|=\epsilon_2}} \, \omega
\end{equation}
does not depend on the $\epsilon_i$ and is called the \textit{Grothendieck residue} of $\omega$ in $0$. Of course, it can be extended to any $z_0 \in \mathbb{C}^n$ simply by making the change of variables $z = z_0 + z'$. The subsets $D_i:=\{z\in\mathbb{C}^n, f_i(z) =0 \}$ for $i=1,2$ are called the \textit{divisors} of $\omega$. In particular, if $f_1(z)=z_1 , \dots f_n(z)=z_n$ then, sequentially applying the one-dimensional Cauchy formla gives
\begin{equation}\label{Cauchy_0}
 \res_0 \, \omega  = \, \frac{1}{(2i\pi)^n} \int\limits_{\substack{|z_1|=\epsilon_1\\ \dots \\ |z_n|=\epsilon_n}} g \, \frac{dz_1}{z_1} \wedge \dots \wedge \frac{dz_n}{z_n}
 = g(0_{\mathbb{C}^n})
\end{equation}
which is known as the \textit{Cauchy formula on the polydisks}.

\textbf{\underline{Example.}} consider the 2-form $\omega = \Gamma(az_1) \Gamma(bz_2) d{z}$ for $a,b\in \mathbb{C}$; $\omega$ is singular in $\underline{z}=(-\frac{n}{a},-\frac{m}{b})$ for all $n,m \in \mathbb{N}$ and we know from the singular series for the Gamma function (\ref{singular_series_gamma}) that around these singularities $\omega$ admits the behavior
\begin{equation}\label{singular_exemple}
\omega \underset{z\rightarrow(-\frac{n}{a},-\frac{m}{b})}{\sim} \frac{(-1)^n}{n!} \frac{(-1)^m}{m!} \frac{dz_1}{n+a z_1} \wedge\frac{dz_2}{m+a z_2}
\end{equation}
Changing the variables:
\begin{equation}
\tilde{z_1}=n+az_1 \,\, , \,\, \tilde{z_2}=m+az_2\,\, , \,\, d\tilde{z_1} \wedge d\tilde{z_2} = ab \, dz_1 \wedge dz_2
\end{equation}
brings back the residue computation in $(0,0)$; applying (\ref{Cauchy_0}) gives
\begin{equation}
\res_{(-\frac{n}{a},-\frac{m}{b})} \omega \, = \, \frac{1}{ab} \frac{(-1)^{n+m}}{n!m!}
\end{equation}

\subsection{Multidimensional Jordan lemma}
\label{MultiJordan}

We denote by $\langle \, . \, | \, . \, \rangle$ the Euclidean scalar product in $\mathbb{C}^n$. Let $x=(x_1, \dots , x_n)$ a vector in $\mathbb{R}^n$ and $\omega$ be the complex differential $n$-form:
\begin{equation}
\omega \, = \, \frac{\prod\limits_{j=1}^{N_j}\Gamma (\langle a_j | z \rangle + b_j) }{\prod\limits_{k=1}^{N_k}\Gamma (\langle c_k | z \rangle + d_k)} \, x_1^{-z_1} \, \dots \, x_n^{-z_n} \, dz\,\, , \,\,\,\,\, a_j , c_k \in \mathbb{R}^n \,\, , \,\,\, b_j,d_k \in \mathbb{R}
\end{equation}
Let $\gamma = (\gamma_1 , \dots , \gamma_n)\in\mathbb{R}^n$ and denote
\begin{equation}
\gamma \, + \, i \mathbb{R}^n \, = \, (\gamma_1-i\infty , \gamma_1+i\infty) \, \times \, \dots \, \times \, (\gamma_n-i\infty , \gamma_n+i\infty)
\end{equation}
We want to know whether the $n$-dimensional Mellin-Barnes integral
\begin{equation}\label{MBn}
\Phi (x) \, = \, \frac{1}{(2i\pi)^n} \int\limits_{\gamma + i \mathbb{R}^n} \, \omega
\end{equation}
can be expressed as a sum of Grothendieck residues, and, if it is the case, in which region of $\mathbb{C}^n$. To that extent, we introduce the $n$-dimensional analogues to (\ref{Delta_1}) and (\ref{Pi_Delta_1}):
\begin{equation}\label{Delta_n}
\Delta \, =: \, \sum\limits_{j=1}^{N_j} a_j \, - \, \sum\limits_{k=1}^{N_k} c_k
\end{equation}
and
\begin{equation}\label{Pi_Delta_n}
\Pi_\Delta \, = \, \left\{ z\in\mathbb{C}^n \, , \,\, Re \langle \Delta | z \rangle \, < \, \langle \Delta \, , \, \gamma  \rangle    \right\}
\end{equation}
and introduce the concept of \textit{compatibility} \cite{Tsikh94,Tsikh97}.

\subsubsection{Compatibility of cones in $\mathbb{C}^n$}

Let $c_j\in\mathbb{R}$ and $h_j:\mathbb{C}\rightarrow\mathbb{C}$ a linear application; let $\Pi_j$ be the subset of $\mathbb{C}$ defined by
\begin{equation}\label{Pi_j}
\Pi_j \, := \, \left\{ z_j\in\mathbb{C},\, Re \left( h_j(z_j) \right) \leq c_j  \right\}
\end{equation}
A \textit{cone} in $\mathbb{C}^n$ is a subset of the type
\begin{equation}
\Pi \, = \, \Pi_1 \, \times \, \dots \, \times \, \Pi_n
\end{equation}
where all the $\Pi_j$ are of type (\ref{Pi_j}); its \textit{faces} $\varphi_j$ are
\begin{equation}
\varphi_j \, := \, \partial \Pi_j
\end{equation}
and its \textit{distinguished boundary}, or \textit{vertex}, is
\begin{equation}
\partial_0 \Pi \, := \, \varphi_1 \, \cap \, \dots \, \cap \varphi_n
\end{equation}
Let $D_k$, $k=1 \dots n$ be the divisors of the $n$-form $\omega$ in (\ref{MBn}), and $D=\cup_{k=1}^n D_j$. A cone $\Pi$ is said to be \textit{compatible} with $D$ if
\begin{itemize}
\item Its distinguished boundary is $\gamma$;
\item Any $D_k$ intersect at most one of its faces $\varphi_j$:
\begin{equation}
D_k \, \cap \, \varphi_j \, = \, \emptyset \,\,\, \textrm{if} \,\,\, i \neq j
\end{equation}
\end{itemize}

\subsubsection{Residue theorem for multiple Mellin-Barnes integrals}

It is shown in \cite{Tsikh94} that,if $\Delta \neq 0$, a Jordan condition is automatically satisfied for the $n$-form $\Omega$ in every compatible cone in $\Pi_\Delta$, that is, the growth of the Gamma can be uniformly controlled and be shown to vanish when $|z|\rightarrow \infty$. In \cite{Tsikh97}, the authors extend their results to the case $\Delta = 0$. The integral over $\gamma + i\mathbb{R}^n$ therefore resumes to a sum of residues (in the sense of Grothendieck) in the compatible cone:
\begin{center}
\fbox{
\begin{minipage}[b]{\textwidth}
\begin{itemize}
\item If $\Delta \neq 0$, then there exists a compatible cone $\Pi\subset \Pi_\Delta$ into which:
\begin{equation}\label{Residue_thm_n}
\frac{1}{(2i\pi)^n} \int\limits_{\gamma + i \mathbb{R}^n} \, \omega \, = \, \sum\limits_{z_k\in\Pi} \, \res _{z_k} \omega
\end{equation}
\item If $\Delta = 0$, the residue summation (\ref{Residue_thm_n}) holds in every compatible cone $\Pi$.
\end{itemize}
\end{minipage}
}
\end{center}
\textbf{\underline{Example.}} Let $f(x_1,x_2)=e^{-(x_1+x_2)}$; introducing sequentially the Mellin pair $e^{-x}\leftrightarrow \Gamma(z)$ which holds with fundamental strip $\langle 0 , \infty \rangle$, we obtain the representation
\begin{equation}
f(x_1,x_2) \, = \, \frac{1}{(2i\pi)^2} \int\limits_{\gamma + i \mathbb{R}^2} \, \Gamma(z_1) \, \Gamma(z_2) \, x_1^{-z_1}x_2^{-z_2} dz \,\,\,\, , \,\,\,\, \gamma_1,\gamma_2 \in (0,\infty) \, \, , \,\, \gamma = \,
\begin{bmatrix}
\gamma_1 \\
\gamma_2
\end{bmatrix}
\end{equation}
The resulting fundamental "polyhedra" $\langle 0 \, , \, \infty \rangle ^2$
is the direct product of the fundamental strips in each Mellin variable. From the singular behavior of the Gamma function, it is clear that $\omega$ possesses two divisors which turn out to be the two families
 \begin{equation}
D_1 \, := \, \{ z_1=-n \,\, , \,\, n\in\mathbb{N} \} \,\,\, , \,\,\, D_2 \, := \, \{ z_2=-m  \,\, , \,\, m\in\mathbb{N} \}
\end{equation}
Computing the characteristic $\Delta$ from definition (\ref{Delta_n}):
\begin{equation}
\Delta = \,
\begin{bmatrix}
1 \\
1
\end{bmatrix}
\end{equation}
and consequently, from definition (\ref{Pi_Delta_n}) the half-plane $\Pi_\Delta$ is
\begin{equation}
\Pi_\Delta \, = \, \{  z\in\mathbb{C}^2 \, , \,\, z_1 + z_2 \, < \, \gamma_1 + \gamma_2 \}
\end{equation}
which, in the $(Re(z_1),Re(z_2))$-plane, is the half-plane located under the line $l_\Delta: z_2 = -z_1 + \gamma_1 + \gamma_2$. If we choose $\Pi$ to be the cone
\begin{equation}
\Pi \, = \, \{ Re(z_1) \, < \, \gamma_1  \} \, \times \, \{ Re(z_2) \, < \, \gamma_2  \}
\end{equation}
it is clear, as shown in Fig. 1, that $\Pi$ is compatible with $D_1\cup D_2$.

\begin{figure}[!h]
\centering
\includegraphics[scale=0.5]{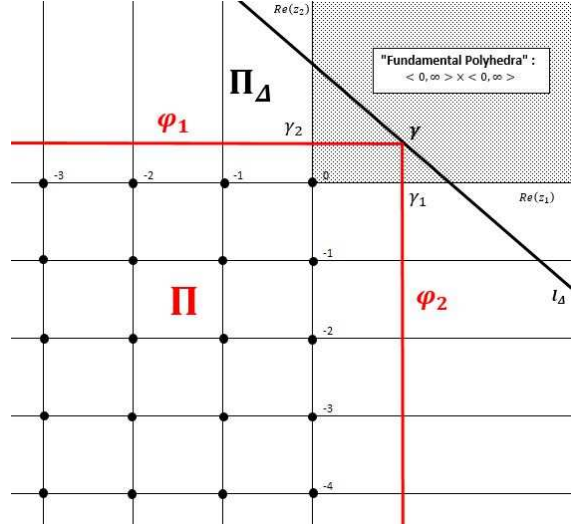}
\caption{The cone $\Pi$, located in the half-plane $\Pi_\Delta$, is compatible with the family of divisors $D_1$ (vertical lines, which only intersect the face $\varphi_1$) and $D_2$ (horizontal lines, which only intersect the face $\varphi_2$). Dotted points correspond to the intersection family $D_1\cap D_2$ and give birth to a residue whose sum  correspond to the integral of $\omega$ over $\gamma + i \mathbb{R}^2$ (\ref{Residue_thm_n}).}
\end{figure}

From the singular series of the Gamma function, it is clear that $\omega$ possesses a (Grothendieck) residue at each point $(-n,-m)$, around which it has the behavior
\begin{equation}
\omega \underset{(z_1,z_2)\rightarrow(-n,-m)}{\sim} \frac{(1)^{n+m}}{n!m!}\frac{x^{-z_1}dz_1}{z_1+n} \, \wedge \, \frac{x^{-z_2}dz_1}{z_2+n}
\end{equation}
From the residue theorem (\ref{Residue_thm_n}) and the Cauchy formula (\ref{Cauchy_0}):
\begin{equation}
e^{-(x_1+x_2)} \, = \, \sum\limits_{z_k\in\Pi} \, \res_{z_k} \, \omega \, = \, \sum\limits_{n,m \in \mathbb{N} } \, \frac{(1)^{n+m}}{n!m!} x_1^n x_2^m
\end{equation}
as expected from the usual Taylor expansion for $e^{-u}$.

\subsection{Series formula for an European Black-Scholes call}

We come back to the evaluation of the European call under Black-Scholes model; from (\ref{BS_Green3}), we introduce the complex differential $2$-form
\begin{equation}
\omega_{BS} \, := \, 2^{\frac{t_2-t_1}{2}}
N(t_1,t_2)
\Gamma(-\frac{t_1}{2})\Gamma(\frac{t_2}{2}) (\sigma\sqrt{\tau})^{-(1+t_1+t_2)} \, dt_1 \, \wedge \, dt_2
\end{equation}
so as to write
\begin{equation}\label{BS_form}
V_{BS} (S,K,\tau) \, = \, \frac{Ke^{-r\tau}}{4\sqrt{2\pi}} \, \frac{1}{(2i\pi)^2} \, \int\limits_{\gamma + i\mathbb{R}^2} \, \omega_{BS}
\end{equation}
Introducing the notation
\begin{equation}\label{[log]}
[\log] \, := \, \log \frac{S}{K} \, + \, r\tau
\end{equation}
and using the decomposition $\frac{1}{s(1+2i\pi s)} \, = \, \frac{1}{s}-\frac{2i\pi}{1+2i\pi s}$ we can write:
\begin{multline}
N(t_1,t_2) \, = \, (-2i\pi)^{-(1+\frac{t_2}{2})} \times \\
\left[
\underbrace{\left\langle s^{-(1+\frac{t_2}{2})} \Omega_{t_1}(-2i\pi s) \, , \, e^{-2i\pi s[\log]} \, \right\rangle}_{:= \, N^{(0)}(t_1,t_2)}
- 2 i \pi \underbrace{\left\langle \frac{s^{-\frac{t_2}{2}}}{1+2i\pi s} \Omega_{t_1}(-2i\pi s) \, , \, e^{-2i\pi s[\log]} \, \right\rangle}_{:= \, N^{(-1)}(t_1,t_2)}
\right]
\end{multline}
Making the change of variables $s=s'-\frac{1}{2i\pi}$ in $N^{(-1)}(t_1,t_2)$ immediately shows that
\begin{equation}
2i\pi N^{(-1)}(t_1,t_2) \, = \, N^{(0)}(t_1,t_2) \, \times \, (-1)^{-\frac{t_2}{2}} \, \times \, e^{[\log]}
\end{equation}
Using the definition (\ref{[log]}) we thus have:
\begin{equation}\label{N0-1}
N(t_1,t_2) \, = \, N^{(0)}(t_1,t_2) \, \left( 1 - (-1)^{-\frac{t_2}{2}}\frac{S}{K}e^{r\tau} \right)
\end{equation}
The residues of $\omega_{BS}$, and therefore the integral (\ref{BS_form}), are thus determined by the behavior of the $N^{(0)}(t_1,t_2)$ kernel around the singularities of $\omega_{BS}$. From the presence of the $\Gamma(-\frac{t_1}{2})$ and $\Gamma(\frac{t_2}{2})$ terms, the divisors of $\omega_{BS}$ are:
\begin{equation}
D_1:=\{ z_1=+2n \, , \,\, n \in\mathbb{N} \} \,\, , \,\, D_2:=\{ z_2=-2m \, , \,\, m \in\mathbb{N} \} \,\, , \,\, D:=D_1 \cup D_2
\end{equation}
The characteristic quantity:
\begin{equation}
\Delta \, = \,
\begin{bmatrix}
-\frac{1}{2} \\ \frac{1}{2}
\end{bmatrix}
\end{equation}
shows that $\Pi_\Delta$ is the half-plane located under the line $l_\Delta \, : \, z_2 = z_1 - \gamma_1 + \gamma_2$.

\begin{figure}[!h]
\centering
\includegraphics[scale=0.7]{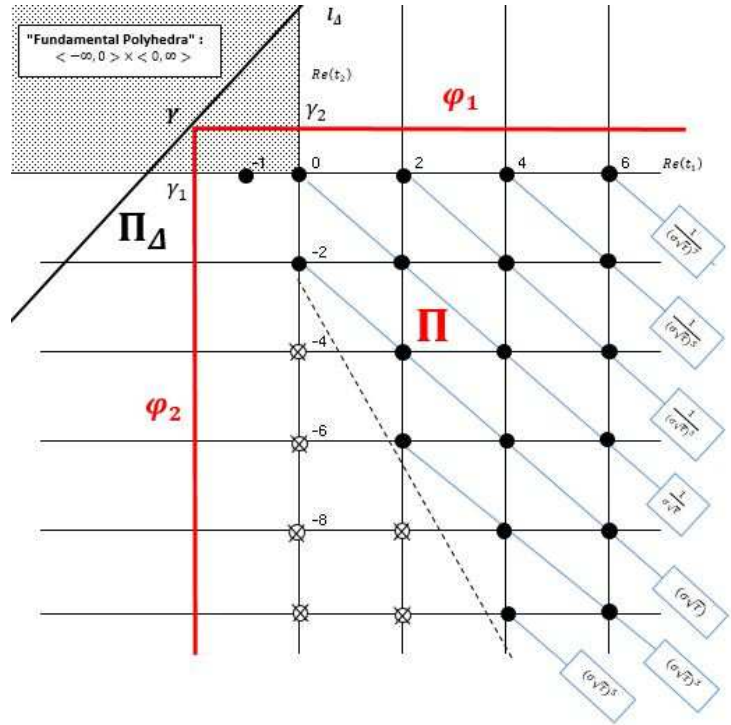}
\caption{The cone $\Pi$ is compatible with the divisors family $D_1$ (vertical lines) and $D_2$ (horizontal lines) of $\omega_{BS}$. The residues arise from an isolated singularity (which gives birth to the "forward" term), and from a sum of terms which can be seen as a "corrective" series in odd powers of $\sigma\sqrt{\tau}$. Among these singularities, located in $(2n,-2m), \, n,m\in\mathbb{N}$ , only those satisfying $1+2n-m$ (above the dotted line) have a non-zero residue and thus contribute to the series.}
\end{figure}

\noindent In the half-plane $\Pi_\Delta$, singularities arise at $t=(2n,-2m), \, n,m \in\mathbb{N}$, because we know from the singular series (\ref{singular_series_gamma}) that
\begin{equation}\label{omega_nm}
\omega_{BS} \, \underset{t \rightarrow (2n,-2m)}{\sim} \, \frac{(-1)^{n+m}}{n!m!} \, 2^{\frac{t_2-t_1}{2}}
N(t_1,t_2)
(\sigma\sqrt{\tau})^{-(1+t_1+t_2)} \, \frac{2dt_1}{2n+t_1} \, \wedge \, \frac{2dt_2}{2m-t_2}
\end{equation}
Now, as $n$ is an integer, it follows from Table 2 that the $\Omega_{2n}(-2i\pi s)$ distribution coincides with derivatives of the Dirac distribution, that is:
\begin{equation}
N^{(0)}(2n,-2m) \, = \, (-1)^{m-1}(2i\pi)^{-2n + m - 1} \left\langle s^{m-1} \delta_{(s)}^{(2n)} \, , \,  e^{-2i\pi s [\log]} \right\rangle
\end{equation}
Using well-known properties of the Dirac distribution
\footnote{Integrating by parts yields (see \cite{Poularikas99} or any usual monograph on Fourier transform):
\begin{align*}
s^N \delta_{(s)}^{(M)} \, = 
\left\{
\begin{aligned}
& (-1)^N \frac{M!}{(M-N)!} \delta_{(s)}^{(M-N)} \,\,\, \textrm{for} \,\,\, M\geq N \in \mathbb{N} \\
& 0 \,\,\, \textrm{for} \,\,\,  M < N
\end{aligned}
\right.
\end{align*}
}, one immediately concludes that
\begin{equation}
N^{(0)}(2n,-2m) \, = \,
\left\{
\begin{aligned}
& \frac{(2n)!}{(1+2n - m)!}[\log]^{1+2n+m} \,\,\ \textrm{ if  } 1+2n - m \geq 0 \\
& 0 \,\,\ \textrm{ if  } 1+2n - m < 0
\end{aligned}
\right.
\end{equation}
and therefore, using and (\ref{N0-1}) and plugging in (\ref{omega_nm}) we easily obtain, from the Cauchy formula:
\begin{multline}\label{residue_n_m}
\res_{(2n,-2m)} \omega_{BS} \, = \, 4(-1)^{n+m} \, \times \\
 \frac{(2n)!}{2^{n+m}n!m!(1+2n-m)!} \left(-1+(-1)^m \frac{S}{K} e^{r\tau} \right) [\log]^{1+2n+m}(\sigma\sqrt{\tau})^{-(1+2n-2m)}
\end{multline}
on the condition that $1+2n-m \geq 0$. Before concluding, one must observe that, in addition of the $D_1 \cup D_2 $-induced singularities, there is one "isolated" singularity induced by the $N(t_1,t_2)$ kernel itself: because
\begin{equation}
N^{(0)}(t_1,0) \, = \, -\left\langle \frac{\Omega_{t_1}(-2i\pi s)}{2i\pi s} \, , \, e^{-2i\pi s [\log]}  \right\rangle
\end{equation}
From the singular series (\ref{singular_series_omega}) we know that
\begin{equation}
N(t_1, 0) \underset{t_1 \rightarrow -1}{\sim} \, = \, \frac{1}{1+t_1} \left\langle \delta_{(s)} \, , \, e^{-2i\pi s [\log]}  \right\rangle \, = \, \frac{1}{1+t_1}
\end{equation}
Therefore
\begin{equation}
\omega_{(BS)} \, \underset{t \rightarrow (-1,0)}{\sim} \,
2^{\frac{t_2-t_1}{2}}
\Gamma(-\frac{t_1}{2})
(\sigma\sqrt{\tau})^{-(1+t_1+t_2)} \, \left( 1 - \frac{S}{K}e^{r\tau} \right) \, \frac{dt_1}{1+t_1} \, \wedge \, \frac{2dt_2}{-t_2}
\end{equation}
Using $\Gamma(\frac{1}{2})=\sqrt{\pi}$, it follows immediately that
\begin{equation}\label{res_-1_0}
\res_{(-1,0)} \omega_{(BS)} \, = \, 2\sqrt{2\pi} \left( -1 + \frac{S}{K}e^{r\tau} \right)
\end{equation}
The singularities in $(2n,-2m)$ and the isolated one are all located in the cone
\begin{equation}
\Pi:=\left\{Re z_1 > \gamma_1 \right\} \, \times \, \left\{Re z_2 < \gamma_2 \right\} \,\subset\Pi_{\Delta}
\end{equation}
which is compatible with $D_1$ and $D_2$. Using the residue theorem (\ref{Residue_thm_n}) we have
\begin{equation}
V_{BS}(S,K,\tau) \, = \, \frac{Ke^{-r\tau}}{4\sqrt{2\pi}} \sum\limits_{t_k\in\Pi} \res_{t_k} \omega_{BS}
\end{equation}
Using the residues formulas (\ref{residue_n_m}) and (\ref{res_-1_0}) we obtain:

\begin{center}
\fbox{
\begin{minipage}[b]{\textwidth}
\textbf{\underline{Series Representation of the Black-Scholes formula}} \\
Let $[\log]=\log\frac{S}{K} + r\tau$, then the Black-Scholes price of an European call option can be decomposed as:
\begin{align}\label{series_BS}
V_{BS}(S,K,\tau) & = \, \frac{1}{2}(S-Ke^{-r\tau})  \nonumber \\
& + \, \frac{1}{\sqrt{2\pi}} \sum\limits_{1+2n-m \geq 0} \,
\frac{(-1)^n(2n)!}{2^{n+m}n!m!(1+2n-m)!}\left(S-(-1)^m Ke^{-r\tau} \right)  \nonumber \\
& \hspace*{5cm} \times[\log]^{1+2n-m} (\sigma\sqrt{\tau})^{-1+2(m-n)}
\end{align}
\end{minipage}
}
\end{center}

\noindent Formula (\ref{series_BS}) is a particular case of a generic formula obtained in \cite{AC16} for a Levy-stable distribution with stability $\alpha$ (the Black-Scholes model corresponding to the specific case $\alpha=2$).

\textbf{\underline{Discussion}}

\noindent Formula (\ref{series_BS}) can be interpreted as a sum of a "forward term", i.e. the difference between the spot price $S$ and the discounted value of the strike at maturity $Ke^{-r\tau}$, and a corrective series that takes into account the market volatility $\sigma$. The fact that only odd powers of $\sigma\sqrt{\tau}$ are concerned is no surprise if one remembers the series expansion of the error function
\begin{equation}
\mathrm{erf}(x) \, = \, \frac{2}{\sqrt{\pi}} \, \left( x - \frac{1}{3}x^3 + \frac{1}{10}x^5 - \frac{1}{42}x^7 + \dots   \right) \,
\end{equation}
which is the basis of the cumulative distribution function $N(u)=\frac{1}{2}\left[ 1 + \mathrm{erf}\left( \frac{u}{\sqrt{2}} \right)  \right]$ in the usual Black-Scholes formula
\begin{equation}
V_{BS}(S,K,\tau) \, = \, S N\left( \frac{1}{\sigma\sqrt{\tau}}[\log] +\frac{1}{2}\sigma\sqrt{\tau} \right) \, - \, K N\left( \frac{1}{\sigma\sqrt{\tau}}[\log] -\frac{1}{2}\sigma\sqrt{\tau} \right)
\end{equation}
Making some numerical tests shows that in fact the first few terms in the series (\ref{series_BS}) already give a satisfying approximation of the option price. Choosing $S=3700$, $K=4000$, $\tau=1$, $\sigma=25\%$ and $r=1\%$, the Black-Scholes formula gives a call price of $264.82$ while
\begin{align}
 & \frac{1}{2}(S-Ke^{-r\tau}) \nonumber \\
+ & \frac{1}{\sqrt{2\pi}}\left(\frac{1}{2}(S+Ke^{-r\tau})-\frac{1}{8}(S-Ke^{-r\tau})[\log]\right)(\sigma\sqrt{\tau}) \nonumber  \\
+ & \frac{1}{\sqrt{2\pi}} \left(  (S-Ke^{-r\tau})[\log] - \frac{1}{4}(S+Ke^{-r\tau})[\log]^2 \right) \frac{1}{\sigma\sqrt{\tau}} \nonumber  \\
- & \frac{1}{\sqrt{2\pi}} \frac{1}{48} (S+Ke^{-r\tau}) (\sigma\sqrt{\tau})^3 \nonumber \\
= &  264.79
\end{align}

\section{Evaluation of Laplace integrals}
\label{sec:EvalLaplace}

This section is devoted to applying the $\mathbb{C}^n$ theory we have developed in the former sections (with its distributional aspects) to the computation of as class of Laplace integrals:
\begin{equation}\label{Laplace}
\frac{1}{2i\pi}\int\limits_{\mu-i\infty}^{\mu + i\infty } e^{px} \, \varphi(p) \, dp
\end{equation}
where we will be particularly interested to the case where $\varphi(p)$ are irrational algebraic functions, which turn out to be crucial in the evaluation of American options.

\subsection{Schwinger's parameter and distributional consequences}
Let $\nu\in\mathbb{R}$
Let $\alpha,\nu \in \mathbb{R}$. It is a direct consequence of the properties in Table 1 that
\begin{equation}\label{Schwinger}
\frac{1}{(p+\alpha)^\nu} \, = \, \frac{1}{\Gamma(\nu)} \, \int\limits_0^\infty \, e^{-(p+\alpha)y} \, y^{\nu -1}  \, dy
\end{equation}
Observation (\ref{Schwinger}) is known as the \textit{Schwinger's trick} and was historically introduce to simplify the computation of loop integrals in Quantum Field Theory; it also greatly simplifies the calculation if several Laplace integrals.

\textbf{\underline{Example 1}} Consider $\varphi(p)=\frac{1}{p^\nu}$ in (\ref{Laplace}) and use the Schwinger trick with $\alpha=0$:
\begin{align}
f_0(x) & = \,  \frac{1}{2i\pi}\int\limits_{\mu-i\infty}^{\mu + i\infty } \frac{e^{px}}{p^\nu} \,  dp \, = \, \frac{1}{2i\pi}\int\limits_{\mu-i\infty}^{\mu + i\infty }  \frac{1}{\Gamma(\nu)} \, \int\limits_0^\infty e^{p(x-y)} y^{\nu -1} \, dy \, dp
\end{align}
Performing the change of variables $p=2i\pi p'$ in the $p$-integral yields
\begin{equation}\label{f_0_1}
f_0(x) \, = \, \int\limits_0^\infty \frac{1}{\Gamma(\nu)} \int\limits_{-\infty}^{+\infty } e^{2i\pi p' (x-y)} \, dp' \, y^{\nu - 1} \, dy
\end{equation}
But, with the notations of section 1, we recognize
\begin{equation}
\int\limits_{-\infty}^{+\infty } e^{2i\pi p' (x-y)} \, dp' \, = \, \Omega_0 \left( 2i\pi (x-y) \right) \, = \, \delta_{(x-y)}
\end{equation}
Replacing in (\ref{f_0_1}), we get
\begin{equation}\label{f_0}
f_0(x) \, = \, \frac{1}{\Gamma(\nu)} \left\langle \delta_{(x-y)} \, , \, y^{\nu-1} \right\rangle \, = \, \frac{x^{\nu-1}}{\Gamma(\nu)}
\end{equation}
A particular case of $(\ref{f_0})$: for $\nu=\frac{1}{2}$, then $\Gamma(\nu) = \sqrt{\pi}$ and therefore
\begin{equation}
\frac{1}{2i\pi}\int\limits_{\mu-i\infty}^{\mu + i\infty } \frac{e^{px}}{\sqrt{p}}\, dp \, = \, \frac{1}{\sqrt{\pi x}}
\end{equation}

\textbf{\underline{Example 2 (Generalization)}} Let $n\in\mathbb{N}$ and $\varphi(p)=\frac{p^n}{(p+\alpha)^\nu}$ in (\ref{Laplace}); and let
\begin{equation}\label{f_n_alpha}
f_{n,\alpha,\nu}(x) \, = \, \frac{1}{2i\pi} \int\limits_{\mu-i\infty}^{\mu + i\infty } e^{px} \frac{p^n}{(p+\alpha)^\nu}\, dp
\end{equation}
Using Schwinger's trick and the change of variables $p=2i\pi p'$ yields
\begin{equation}\label{f_n_alpha_1}
f_{n,\alpha,\nu}(x) \, = \, \frac{(2i\pi)^n}{\Gamma(\nu)} \int\limits_0^\infty \int\limits_{-\infty}^{+\infty } e^{2i\pi p' (x-y)} \, p'^n \, dp' \, e^{-\alpha y } \, y^{\nu - 1} \, dy
\end{equation}
But, as
\begin{equation}
 \int\limits_{-\infty}^{+\infty } e^{2i\pi (x-y)} \, p'^n \, dp' \, = \, \Omega_{n}(2i\pi '(x-y)) \, = \, \frac{1}{(2i\pi)^n} \, \delta_{(x-y)}^{(n)}
\end{equation}
Replacing in (\ref{f_n_alpha_1}) yields
\begin{equation}\label{f_n_alpha_2}
f_{n,\alpha,\nu}(x) \, = \, \frac{1}{\Gamma(\nu)} \left\langle \delta_{(x-y)}^{(n)} \, , \, e^{-\alpha y } \, y^{\nu - 1} \right\rangle
\end{equation}
We can introduce the \textit{generalized Laguerre polynomials}
\begin{equation}\label{Laguerre}
\left\langle \delta_{(x-y)}^{(n)} \, , \, e^{-\alpha y } \, y^{\nu} \right\rangle \, := \, L_n^{(\alpha)}(x,\nu) \, := \, e^{-\alpha x}l_n^{(\alpha)}(x,\nu)
\end{equation}
The first polynomials are:
\begin{align}\label{Laguerre_examples}
\left\{
\begin{aligned}
& l_0^{(\alpha)}(x,\nu) \, = \, x^\nu \\
& l_1^{(\alpha)}(x,\nu) \, = \, \alpha x^{\nu} - \nu x^{\nu-1} \\
& l_2^{(\alpha)}(x,\nu) \, = \, -\alpha^2 x^\nu + 2\alpha\nu x^{\nu-1} - \nu(\nu-1) x^{\nu-2} \\
& \dots
\end{aligned}
\right.
\end{align}
In particular for $n=0$, $\alpha=0$, $\nu=\frac{1}{2}$ one has
\begin{equation}
\frac{1}{2i\pi}\int\limits_{\mu-i\infty}^{\mu + i\infty }  \frac{e^{px}}{\sqrt{1+p}}\, dp \, = \, \frac{e^{-x}}{\sqrt{\pi x}}
\end{equation}

\subsection{The American option kernel}
\label{AmericanOptioKernel}
The evaluation of American options is a more complicated matter than European option pricing, even in the simple Black-Scholes framework. The chief difference is the presence of an optimal exercise price $S_f(\tau)$ which converts the Black-Scholes PDE with terminal condition into a \textit{free-boundary}, or \textit{Stefan} problem. Estimating this optimal price is crucial as it allows to bring back the problem to a classical European fixed-boundary one; several efforts have been made in that sense, notably into the form of analytic-approximation formulas for $S_f(\tau)$ \cite{Chen00,Bunch00} but it turns out that these approximation formulas are only valid for small $\tau$, or call for an intermediate numerical resolution of a non-linear equation. A very interesting and more generic approach is given in \cite{Zhu06}: the author solves the free boundary Black-Scholes system in the Laplace space and, assuming a pseudo steady-state approximation when performing the Laplace transform, obtains the remarkable representation for the optimal exercise price:
\begin{equation}\label{Sf_Laplace}
S_f(\tau) \, = \, \frac{1}{2i\pi} \int\limits_{\mu-i\infty}^{\mu+i\infty} \frac{e^{p\tau}}{p} \, \textrm{exp} \left\{ \frac{-\log\left[1-\frac{p+\gamma}{\gamma(b-\sqrt{p+a^2})}\right]}{b+\sqrt{p+a^2}} \right\} \, dp \,\,\ , \,\,\,\, \mu>0 
\end{equation}
where the constants are related to the riskless interest rate $r$ and the market volatility $\sigma$ by
\begin{equation}
\gamma \, = \, \frac{2r}{\sigma^2} \,\, , \,\, a=\frac{1+\gamma}{2} \,\, , \,\, b=\frac{1-\gamma}{2} \,\ , \,\, a^2 = b^2 + \gamma
\end{equation}
Now, if we introduce suitable Mellin-Barnes representations for $e^{-x}$ and $\log^k(1+x)$ in the Laplace integral (\ref{Sf_Laplace}), we obtain a factorized Mellin-Barnes representation for the optimal exercise price
\begin{equation}\label{Sf_Mellin_1}
S_f(\tau) \, = \, \frac{1}{(2i\pi)^2} \, \int\limits_{c+i\mathbb{R}^2}  \, \mathcal{S}(t_1,t_2) \, A_{-t_1,-t_2} (\tau) \, \gamma^{t_2} \, dt_1 \wedge dt_2 \,\,\, , \,\,\,\, c:=\begin{bmatrix} c_1 \\ c_2 \end{bmatrix}
\end{equation}
where
\begin{itemize}
\item $\mathcal{S}(t_1,t_2)$ possesses Gamma-function induced singularities at every strictly negative integers $t_1 = -n$ and $t_2=-m$.
\item The kernel $A_{-t_1,-t_2} (\tau)$ is analytic for theses values of $(t_1,t_2)$ and is defined by the remaining Laplace integral:
\begin{equation}\label{A_n_m1_m2}
A_{n,m}(\tau) \, := \, \frac{1}{2i\pi}\int\limits_{\mu-i\infty}^{\mu + i\infty} \frac{e^{p\tau}}{p}\frac{(p+\gamma)^m}{(b+\sqrt{p+a^2})^{n}(b-\sqrt{p+a^2})^{m}} \, dp
\end{equation}
\end{itemize}
The factorized representation (\ref{Sf_Mellin_1}) is extremely interesting because it shows that the value of $S_f(\tau)$ can be analytically expressed as a residue series in power of $\gamma$; these powers are driven by the singular behavior of $\mathcal{S}(t_1,t_2)$, and their coefficient are determined by the value of the kernel $A_{-t_1,-t_2} (\tau)$ at these singularities. Here, we show how to compute these values; the overall calculation will be detailed in a future publication \cite{ACKK16-American}.

\textbf{\underline{Step 1:}} Write a Schwinger trick for each term in the denominator of the second fraction in (\ref{Sf_Mellin_1}) so as to get:
\begin{multline}\label{A_n_m1_m2_2}
A_{n,m}(\tau) \, = \, \frac{1}{2i\pi} \, \int\limits_{\mu-i\infty}^{\mu + i\infty} \frac{e^{p\tau}(p+\gamma)^m}{p} \, \times  \\  \frac{1}{\Gamma(n)\Gamma(m)} \int\limits_0^\infty \int\limits_0^\infty e^{-(b+\sqrt{p+a^2})t_1}e^{-(b-\sqrt{p+a^2})t_2} t_1^{n-1}t_2^{m-1} dt_1dt_2dp
\end{multline}

\textbf{\underline{Step 2:}} Use the Mellin-Barnes representation combined with the properties in table 1 to write:
\begin{equation}
e^{-\sqrt{p+a^2}t_1}e^{+\sqrt{p+a^2}t_2} \, = \, \frac{1}{(2i\pi)^2} \, \int\limits_{\gamma + i \mathbb{R}^2} (-1)^{-s_2} \frac{\Gamma(s_1)\Gamma(s_2)}{(p+a^2)^{\frac{s_1+s_2}{2}}}t_1^{-s_1} t_2^{-s_2} ds_1 \, \wedge \, ds_2 
\end{equation}
where $\gamma=\begin{bmatrix} \gamma_1 >0 \\ \gamma_2 >0 \end{bmatrix}$. Replacing in (\ref{A_n_m1_m2_2}):
\begin{multline}
A_{n,m}(\tau) \, = \, \frac{1}{(2i\pi)^2} \int\limits_{\gamma + i \mathbb{R}^2} (-1)^{-s_2} \frac{\Gamma(s_1)\Gamma(s_2)}{\Gamma(n)\Gamma(m)}
\int\limits_0^\infty e^{-b t_1}t_1^{n-s_1-1} \, dt_1
\\
\times \int\limits_0^\infty e^{-b t_2}t_1^{m-s_2-1} \, dt_2
\times \frac{1}{2i\pi}
\int\limits_{\mu-i\infty}^{\mu + i\infty} \frac{e^{p\tau}}{p}\frac{(p+\gamma)^m}{(p+a^2)^{\frac{s_1+s_2}{2}}} \, dp
\end{multline}
It is easy to perform the $t_1$ and $t_2$ integrals and obtain:
\begin{multline}\label{A_n_m1_m2_3}
A_{n,m}(\tau) \, = \, \\
\frac{1}{(2i\pi)^2} \int\limits_{\gamma + i \mathbb{R}^2} \frac{(-1)^{-s_2}}{b^{n+m-s_1-s_2}} \frac{\Gamma(s_1)\Gamma(n-s_1)\Gamma(s_2)\Gamma(m-s_2)}{\Gamma(n)\Gamma(m)} F(\tau,s_1,s_2) \, ds_1 \, \wedge \, ds_2
\end{multline}

\textbf{\underline{Step 3:}} We are now left with a considerably simpler Laplace integral to perform:
\begin{equation}
F(\tau,s_1,s_2) \, := \, \frac{1}{2i\pi}
\int\limits_{\mu-i\infty}^{\mu + i\infty} \frac{e^{p\tau}}{p}\frac{(p+\gamma)^m}{(p+a^2)^{\frac{s_1+s_2}{2}}} \, dp
\end{equation}
Note that from elementary properties of Laplace transform, we know that
\begin{equation}
F(\tau,s_1,s_2) \, = \, \int\limits_{0}^\tau \, f(u, s_1, s_2) \, du
\end{equation}
where
\begin{equation}\label{f(u,s_1,s_2)}
f(u,s_1,s_2) \, = \, \frac{1}{2i\pi}
\int\limits_{\mu-i\infty}^{\mu + i\infty} e^{pu} \frac{(p+\gamma)^m}{(p+a^2)^{\frac{s_1+s_2}{2}}} \, dp
\end{equation}
Perform the change of variables $p=p'-\gamma$ and forgetting the prime, we have:
\begin{equation}
f(u,s_1,s_2) \, = \, e^{-\gamma \tau} \, \frac{1}{2i\pi}
 \, \int\limits_{\mu-i\infty}^{\mu + i\infty} e^{pu} \frac{p^m}{(p+a^2 - \gamma)^{\frac{s_1+s_2}{2}}} \, dp
\end{equation}
We recognize an integral of type (\ref{f_n_alpha}): with formula (\ref{f_n_alpha_2}), we immediately get
\begin{equation}
f(u,s_1,s_2) \, = \frac{1}{\Gamma(\frac{s_1+s_2}{2})} \, \, e^{-\gamma u}  \left\langle \delta_{(u-y)}^{(m)} \, , \, e^{(\gamma-a^2) y } \, y^{\frac{s_1+s_2}{2} - 1} \right\rangle
\end{equation}
It is now simple to integrate by parts to obtain
\begin{equation}
F(\tau,s_1,s_2) \, = \, -\frac{\gamma e^{-\gamma \tau}}{\Gamma(\frac{s_1+s_2}{2})} \left\langle \delta_{(u-y)}^{(m-1)} \, , \, e^{(\gamma-a^2) y } \, y^{\frac{s_1+s_2}{2} - 1} \right\rangle
\end{equation}
Using definition (\ref{Laguerre}), as well as the identity $a^2=b^2+\gamma$, we get
\begin{equation}
F(\tau,s_1,s_2) \, = \, -\frac{\gamma }{\Gamma(\frac{s_1+s_2}{2})} \, e^{-a^2 \tau} \, l_{m-1}^{(b^2)}\left( \tau, \frac{s_1+s_2}{2} -1 \right)
\end{equation}
Consequently, replacing in (\ref{A_n_m1_m2_3}) and denoting $d\underline{s} = ds_1 \wedge ds_2$:
\begin{center}
\fbox{
\begin{minipage}[b]{\textwidth}
\textbf{\underline{Mellin-Barnes representation of the American option kernel}}
\begin{multline}\label{A_n_m1_m2_4}
A_{n,m}(\tau) \, = \,  \frac{1}{(2i\pi)^2} \gamma e^{-a^2 \tau} \\
 \int\limits_{\gamma + i \mathbb{R}^2} \frac{(-1)^{-(1+s_2)}}{b^{n+m-s_1-s_2}} \frac{\Gamma(s_1)\Gamma(n-s_1)\Gamma(s_2)\Gamma(m-s_2)}{\Gamma(n)\Gamma(m)\Gamma(\frac{s_1+s_2}{2})} l_{m-1}^{(b^2)}\left( \tau, \frac{s_1+s_2}{2} -1 \right) \, d\underline{s}
 \end{multline}
\end{minipage}
}
\end{center}

\begin{figure}[!h]
\centering
\includegraphics[scale=0.5]{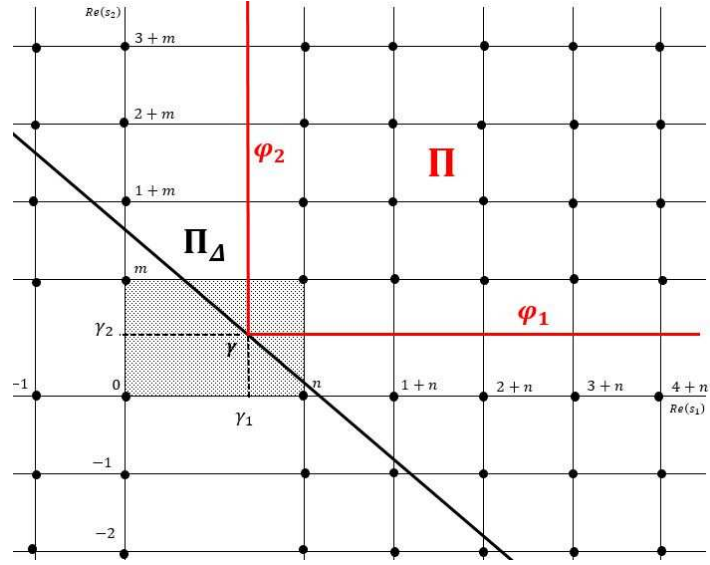}
\caption{The cone $\Pi$ is compatible with the two families of divisors induced by $\Gamma(n-s_1)$ and $\Gamma(m-s_2)$.}
\end{figure}

\noindent We conclude by evaluating the integral (\ref{A_n_m1_m2_4}); denote by $\omega_{n,m}(\tau)$ the integrand in (\ref{A_n_m1_m2_4}) and compute its characteristic quantity
\begin{equation}
\Delta \, = \, \begin{bmatrix}  \frac{1}{2} \\  \frac{1}{2} \end{bmatrix}
\end{equation}
which leads to consider the half-plane $\Pi_{\Delta}$ located over the line $l_\Delta : t_2 = -t_1 +\gamma_1 + \gamma_2$. In this half-plane, the divisors to consider are induced by $\Gamma(n-s_1)$ and $\Gamma(m-s_2)$, that is:
\begin{align}
\left\{
\begin{aligned}
D_1 \, := \, \left\{ s_1 \, = \, k_1 + n \, , \, k_1\in\mathbb{N} \right\} \\
D_2 \, := \, \left\{ s_2 \, = \, k_2 + m \, , \,  k_2\in\mathbb{N} \right\}
 \end{aligned}
\right.
\end{align}
The cone $\Pi$ defined by
\begin{equation}
\Pi:=\left\{Re (s_1) > \gamma_1 \right\} \, \times \, \left\{Re (s_2) > \gamma_2 \right\} \,\subset\Pi_{\Delta}
\end{equation}
is then compatible with $D_1\cup D_2$. The singularities arise in $s_{k_1,k_2}:=(k_1+m_1,k_2+m_2)$ and $\omega$ has the behavior:
\begin{multline}
\omega_{n,m}(\tau) \, \underset{s \rightarrow s_{k_1,k_2}}{\sim} \frac{(-1)^{k_1+k_2}}{k_1!k_2!}\frac{(-1)^{-(1+s_2)}}{b^{n+m-s_1-s_2}} \, \times \\
 \frac{\Gamma(s_1)\Gamma(s_2)}{\Gamma(n)\Gamma(m)\Gamma(\frac{s_1+s_2}{2})} l_{m-1}^{(b^2)}\left( \tau, \frac{s_1+s_2}{2} -1 \right) \, \frac{ds_1}{k_1+n-s_1} \, \wedge \, \frac{ds_2}{k_2+m-s_2}
\end{multline}
Taking the residues and applying (\ref{Residue_thm_n}):
\begin{multline}\label{A_n_m1_m2_series}
A_{n,m}(\tau) \, = \, \frac{\gamma}{\Gamma(n)\Gamma(m)} \, e^{-a^2\tau} \, \times \\
\sum\limits_{k_1,k_2 \in \mathbb{N}} \, b^{k_1+k_2} \, \frac{(-1)^{k_1-m-1}}{k_1! k_2!} \, \frac{\Gamma(k_1+n)\Gamma(k_2+m)}{\Gamma(\frac{k_1+n+k_2+m}{2})} \, l_{m-1}^{(b^2)}\left( \tau, \frac{k_1+n+k_2+m}{2} -1 \right)
\end{multline}

\section{Conclusions}
\label{sec:conclusions}
In this paper, we have discussed some important aspects of Mellin calculus with applications to option pricing. Distributional interpretation of Mellin calculus allows us dealing with Mellin representations of polynomials and logarithmic functions. Inversion formulas for n-dimensional Mellin-Barnes integrals provide a powerful tool for asymptotic analysis of many special functions, including Green functions of fractional diffusion. All these results can be successfully used in financial applications, especially in option-pricing models of European, American and other exotic options. In this paper were presented several examples of options driven by Brownian diffusion and anomalous fractional diffusion, but the possible applications are not limited to these models and can be also used for other types of derivatives. It is also noteworthy to mention that the presented mathematical concepts have a good potential in applications in many scientific field, including quantum mechanics, anomalous diffusion or image processing. Some of these applications will be a subject of the ongoing research.

\end{document}